\documentclass[conference, a4paper]{IEEEtran}

\IEEEoverridecommandlockouts                              % This 
\usepackage{graphicx}
\usepackage{multirow}
\usepackage{amsmath,amssymb,latexsym}
\usepackage{rotating}
\usepackage{lettrine}
\usepackage{textcomp}
\usepackage{units}
\usepackage{hyperref}

\usepackage{xcolor}

\usepackage{lipsum}% For 'Lorem Ipsum' dummy text
\usepackage[pscoord]{eso-pic}% The zero point of the coordinate systemis the lower left corner of the page (the default).
\newcommand{\placetextbox}[3]{% \placetextbox{<horizontal pos>}{<vertical pos>}{<stuff>}
  \setbox0=\hbox{#3}% Put <stuff> in a box
  \AddToShipoutPictureFG*{% Add <stuff> to current page foreground
    \put(\LenToUnit{#1\paperwidth},\LenToUnit{#2\paperheight}){\vtop{{\null}\makebox[0pt][c]{#3}}}%
  }%
}%

\title{Linear Combination of Exponential Moving Averages for Wireless Channel Prediction
\thanks{This work was partially supported by the European Union under the Italian National Recovery and Resilience Plan (NRRP) of NextGenerationEU, partnership on ``Telecommunications of the Future'' (PE00000001 - program ``RESTART'').}
}

\author{
    \IEEEauthorblockN{Gabriele Formis\IEEEauthorrefmark{1}\IEEEauthorrefmark{2}\href{https://orcid.org/0000-0001-9290-002X}{\includegraphics[scale=0.65]{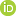}}, Stefano Scanzio\IEEEauthorrefmark{1}\href{https://orcid.org/0000-0001-7643-2342}{\includegraphics[scale=0.65]{orcid_16x16.png}}, Gianluca Cena\IEEEauthorrefmark{1}\href{https://orcid.org/0000-0003-0084-5321}{\includegraphics[scale=0.65]{orcid_16x16.png}}, and Adriano Valenzano\IEEEauthorrefmark{1}\href{https://orcid.org/0000-0003-1238-0808}{\includegraphics[scale=0.65]{orcid_16x16.png}}}
    \IEEEauthorblockA{\IEEEauthorrefmark{1}National Research Council of Italy (CNR--IEIIT), Italy.\\ Email: gabriele.formis@ieiit.cnr.it, stefano.scanzio@cnr.it, gianluca.cena@cnr.it, adriano.valenzano@cnr.it}
    \IEEEauthorblockA{\IEEEauthorrefmark{2}Politecnico di Torino, Italy}
}

\begin{document}
\placetextbox{0.5}{1}{This is the author's version of an article that has been published in this journal.}
\placetextbox{0.5}{0.985}{Changes were made to this version by the publisher prior to publication.}
\placetextbox{0.5}{0.97}{The final version of record is available at \href{https://doi.org/10.1109/INDIN51400.2023.10218083}{https://doi.org/10.1109/INDIN51400.2023.10218083}}%
\placetextbox{0.5}{0.05}{Copyright (c) 2023 IEEE. Personal use is permitted.}
\placetextbox{0.5}{0.035}{For any other purposes, permission must be obtained from the IEEE by emailing pubs-permissions@ieee.org.}%

\maketitle
\thispagestyle{empty}
\pagestyle{empty}

%%%%%%%%%%%%%%%%%%%%%%%%%%%%%%%%%%%%%%%%%%%%%%%%%%%%%%%%%%%%%%%%%%%%%%%%%%%%%%%%
\begin{abstract}
The ability to predict the behavior of a wireless channel in terms of the frame delivery ratio is 
quite valuable, and permits, e.g., 
to optimize the operating parameters of a
wireless network at runtime, or to proactively react to the degradation of the channel quality, 
in order to meet the stringent requirements about dependability and end-to-end latency
that typically characterize
industrial applications.

In this work, prediction models based on the exponential moving average (EMA) are investigated in depth, 
which are proven to outperform other simple statistical methods and whose performance is nearly as good as artificial neural networks, but with dramatically lower computational requirements. 
Regarding the innovation and motivation of this work, a new model that we called EMA linear combination (ELC), is introduced, explained, and evaluated experimentally.
Its prediction accuracy, tested 
on some databases acquired from a real setup based on Wi-Fi devices, showed that ELC brings tangible improvements over EMA in any experimental conditions, the only drawback being a slight increase in computational complexity.
\end{abstract}

%%%%%%%%%%%%%%%%%%%%%%%%%%%%%%%%%%%%%%%%%%%%%%%%%%%%%%%%%%%%%%%%%%%%%%%%%%%%%%%%

\section{Introduction}
\label{sec:introduction}
New paradigms like Industry 4.0 \cite{CANAS2021107379} and beyond \cite{MADDIKUNTA2022100257}
demand for industrial networks with high flexibility and the ability to suitably tackle heterogeneity \cite{SCANZIO2021103388}.
In this respect, wireless communication technologies are one of the main enablers, 
by supporting mobility and automatic (re)configurability of industrial systems \cite{9779183}.
The ever increasing need for wire-free solutions in industries, homes, and cities, 
implies that specific research activities,
aimed at making the related technologies capable to fulfill the very strict requirements, in terms of dependability and end-to-end latency, of industrial networks \cite{9945847, 8502636, LEONARDI202257, 2016-TII-WiRed, 9921559, 10034532},
are more and more demanded.
Besides communication,
services like localization \cite{9137286}, time-synchronization \cite{MONGELLI20161}, and roaming \cite{9149123} 
have to be also provided  
for enabling fully wireless industrial systems
(required, e.g., when devices are characterized by mobility),
by adapting the existing approaches to the specific wireless technology.

If the future quality of a wireless link 
could be predicted in advance with a certain accuracy, 
e.g., in terms of frame delivery ratio (FDR), 
the communication protocol could exploit this information for trying to meet the specific requirements of industrial applications. 
For instance, when the quality of a specific channel is expected to worsen, the application can switch to another channel (suffering from less disturbance) to preserve communication reliability \cite{label} 
or to prevent energy consumption from increasing \cite{electronics11030304}.
As an alternative, it can change some network/communication parameters or diminish the amount of best-effort traffic (shaping) to privilege time-sensitive data exchanges. 
Among the many wireless technologies available today, Wi-Fi and 5G are the most promising for applications with higher demands in terms of dependability and end-to-end latency, while allowing at the same time high throughput. 
This work is based on Wi-Fi, but the technique we present for channel quality prediction can be applied with minimal changes to other wireless communication technologies as well.

Many research works made use of machine learning (ML) for channel quality prediction
in Wi-Fi \cite{9786784}: 
in \cite{9120030} artificial neural networks (ANN) were applied to artificial data; 
in \cite{2022-ITL-ML, 9921698} they were exploited on data derived from real devices;
in \cite{8813020, 9781119562306,8884240} ML is used for the prediction of the channel gain and/or the received signal strength.
The use of ANNs is quite expensive from the computational point of view, and consequently in \cite{WFCS2023} the benefits provided by less CPU-hungry approaches like moving averages and regression models were analyzed in depth.
That work showed that the exponential moving average (EMA) is able to offer the best performance among non-ANN models, and behaved almost as well as an ANN trained only on frame losses. 
Quite interestingly, EMA featured the lowest computational complexity among the analyzed models. This makes it suitable for implementation in any kind of (embedded) device. 

In this paper, an algorithm named EMA linear combination (ELC) is presented and evaluated over a 50-day-long test database, acquired in a real setup that includes several \mbox{Wi-Fi} nodes operating on different channels. The optimal parameters of this algorithm are found by minimizing the mean squared error for the training database.
Results show that ELC provides consistent improvements when compared to the simpler EMA model, yet keeping computational complexity relatively low.
The following Section~\ref{sec:EMA} introduces the concept of channel prediction and the EMA model analytically, and Section~\ref{sec:ELC} describes the new ELC algorithm, which is evaluated in Section~\ref{sec:results}, before the conclusive remarks of Section~\ref{sec:conclusions}.

\section{Channel prediction and EMA model}
\label{sec:EMA}
This work considers a real Wi-Fi link between a STA and an AP.
The wireless channel was probed periodically by means of confirmed one-shot transmissions (no retries allowed) 
with a period $T_s=\unit[0.5]{s}$, corresponding to a frequency of $\unit[2]{Hz}$. 
The receiver node (the AP, in this case) confirms the correct arrival of the $i$-th frame by returning an Acknowledgement (ACK) frame. 
The correct reception of this ACK frame by the sender node is recorded as the outcome $x_i=1$ (\textit{success}). 
Otherwise, if the ACK frame is not received, we logged $x_i=0$ (\textit{failure}). 
The above measurement procedure was practically implemented in real Wi-Fi devices by following the indications reported in Subsection~\ref{sub:DB}.

A database consists of an ordered sequence of outcomes $\mathcal{D}=\{x_1,...,x_i,...,x_{|\mathcal{D}|}\}$,
obtained from the experimental testbed on a time interval long enough (some days).
Two kinds of databases were used in this work:
\textit{training} ($\mathcal{D}_{tr}$), used to train the models by learning some parameters from data, 
and \textit{test} ($\mathcal{D}_{te}$), used to evaluate the proposed models. 
Their sizes, in terms of the number of outcomes, are $|\mathcal{D}_{tr}|$ and $|\mathcal{D}_{te}|$, respectively. 

The goal of this work is finding a simple and effective way to predict the quality of the wireless channel in terms of the FDR, 
i.e., the ratio between the number of ACK frames that come back to the sender node and the number of transmitted frames. 
We relied on the exponentially weighted moving average (EMA) of outcomes $x_i$, which is the simplest form of an infinite impulse response (IIR) low-pass filter, described as
\begin{eqnarray}
\label{eq:EMA}
    y_i & = & \alpha \cdot x_{i} + (1-\alpha) \cdot y_{i-1},
\end{eqnarray}
where $y_i$
is the current FDR prediction, 
$y_{i-1}$ is the previous prediction, 
$x_{i}$ is the current outcome, 
and $\alpha$ is a weight for balancing between present and past.
An $\alpha$ value close to $1$ privileges the current outcome (i.e., the model is more reactive to track sudden changes in channel quality), 
while values close to $0$ prioritize the previous prediction made by EMA (making the output more stable and less susceptible to statistical fluctuations). 
For the very first prediction we selected the value $y_0 = 0.5$, 
but if a better estimation of the channel FDR is available its value should be used to initialize $y_0$.
In the Z-domain, the low-pass filter in \eqref{eq:EMA} is characterized by the transfer function
\begin{align}
    H^{(\alpha)}(z) &= \frac{Y(z)}{X(z)} = \frac{\alpha}{1 - (1-\alpha)z^{-1}}.
\end{align}

\begin{figure}[t]
	\begin{center}
	\includegraphics[width=0.95\columnwidth]{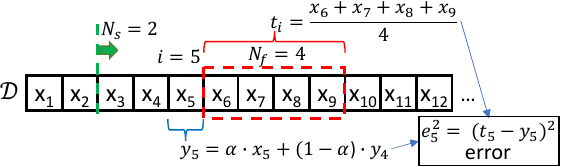}
	\end{center}
	\caption{Example of computation of the EMA model ($y_i$), target ($t_i$), and mean squared error ($e^2_i$) with $i=5$ (the first $N_s$ outcomes are used only for the execution of EMA model, but not for the computation of the error).}
	\label{fig:sliding}
\end{figure}

As highlighted in Fig.~\ref{fig:sliding}, in order to assess the performance of the proposed EMA model
a \textit{target} is computed for every value of $x_i$ in $\mathcal{D}_{te}$, with $1 \leq i \leq |\mathcal{D}_{te}|-N_f$,
starting from the following $N_f$ outcomes ($x_{i+1},...,x_{i+N_f}$) 
\begin{equation}
\label{eq:target}
    t_i = \frac{1}{N_f} \sum_{j=i+1}^{i+N_f}{x_j}.
\end{equation}
This target represents an estimate of the FDR over an interval in the immediate future whose width is 
$T_f={N_f \cdot T_s}$. 
For example, if $N_f=3600$ then $T_f=\unit[30]{min}$,
and the middle point of such interval is located $\unit[15]{min}$ after the current sample.
It is not worth shrinking $N_f$ too much because of statistical fluctuations,
which would make the target unreliable.

The most recent value $y_i$ given by (\ref{eq:EMA}) was taken as the prediction for the target, under the reasonable assumption that,
despite they are displaced by $T_f/2$,
it constitutes the best possible estimate. 
The adoption of predictive models like those based on linear and polynomial regression has been analyzed in \cite{WFCS2023}, 
where their showed poorer prediction accuracy than EMA.

Evaluation of the prediction accuracy was carried out on outcomes with index $i=N_s+1, N_s+2,...,|\mathcal{D}_{te}|$ of $\mathcal{D}_{te}$. 
The first $N_s$ outcomes were discarded because we aim to evaluate steady-state accuracy, neglecting the initial (and potentially long) transient from $y_0$ up to when FDR estimation settles.
The closer $y_i$ and $t_i$ are, the better the prediction ability of the model. 
To quantify the model's accuracy, several kinds of errors were considered: 
the real error $e_i=t_i-y_i$, 
the absolute error $|e_i|=|t_i-y_i|$, and the squared error $e^2=(t_i-y_i)^2$. 
The latter is the one that is minimized by the learning method described in Section~\ref{sec:ELC}.

Starting from these errors, statistical indices can be computed, which include 
the average ($\mu_e$, $\mu_{|e|}$ and $\mu_{e^2}$), 
standard deviation ($\sigma_e$, $\sigma_{|e|}$ and $\sigma_{e^2}$), minimum ($e_{\mathrm{min}}$, ${|e|}_{\mathrm{min}}$ and $e^2_{\mathrm{min}}$), 
percentiles ($e_{p90}$, $e_{p95}$, $e_{p99}$, ${|e|}_{p90}$, ${|e|}_{p95}$, ${|e|}_{p99}$ and $e^2_{p90}$, $e^2_{p95}$, $e^2_{p99}$), 
and maximum ($e_{\mathrm{max}}$, ${|e|}_{\mathrm{max}}$ and $e^2_{\mathrm{max}}$). 
The number of predictions on which these statistics are computed is $|\mathcal{D}_{te}|-N_s-N_f$. 
As an example, the mean squared error (MSE) can be obtained as $\mu_{e^2}=\sum_{i=N_s+1}^{|\mathcal{D}|-N_f}{(t_i-y_i)^2}$.

The EMA model is parameterized by a single quantity, the $\alpha$ weight.
Optimal values can be found for it, each of which minimizes some objective functions for some database. 
We denote $\alpha^*$ the value that minimizes the MSE ($\mu_{e^2}$) for  $\mathcal{D}_{tr}$
\begin{equation}
    \label{eq:ema_opt}
    \alpha^* = \arg \min_{\alpha} \sum_{i=N_s+1}^{|\mathcal{D}_{tr}|-N_f}
\Bigl(t_i - y_i \Bigr)^2.
\end{equation}
To assess in a fair way EMA performance,
the value $\alpha^*$ estimated this way
is then used on the test database $\mathcal{D}_{te}$.

It is worth pointing out that optimality of $\alpha^*$ refers to the whole $\mathcal{D}_{tr}$ ($|\mathcal{D}_{tr}|-N_s-N_f$ values). 
However, larger values for the $\alpha$ weight permit to track FDR variations more quickly, while smaller values 
provide better accuracy when channel conditions are stationary. 
In addition, $\alpha^*$ was estimated on $\mathcal{D}_{tr}$, but it cannot be assured to be optimal for $\mathcal{D}_{te}$ as well.

A mechanism for finding the different $\alpha$ values that represent at best the different FDR variation patterns for a given database $\mathcal{D}$ will be presented in Subsection~\ref{sec:ELC}.

\section{Linear Combination of EMA}

\label{sub:estimation}
\begin{figure*}[t]
    \begin{center}
    \includegraphics[width=2.00\columnwidth]{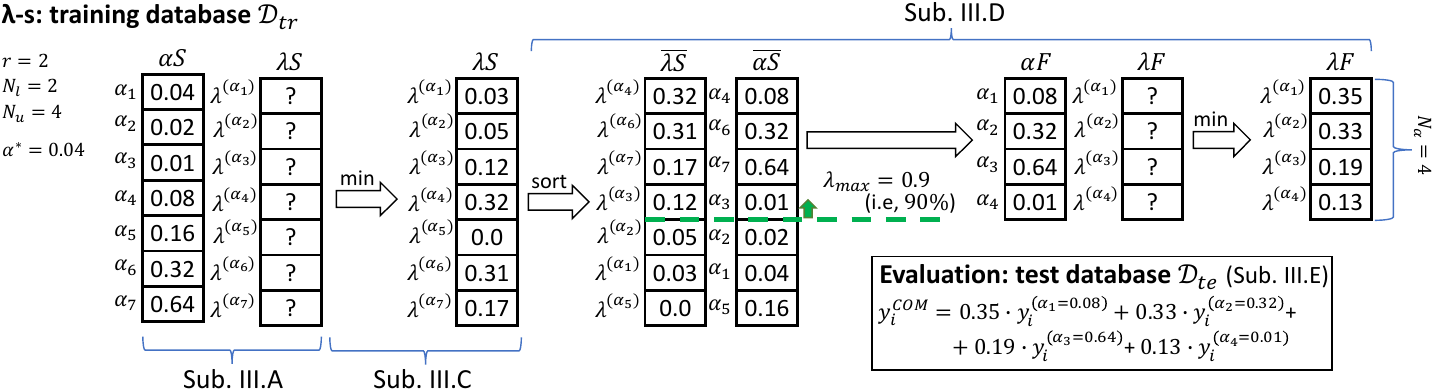}
    \end{center}
    \caption{Example about the application of the ELC algorithm (``min'' refers to the minimization of the linear combination, as explained in Subsection~\ref{sub:minimization}).}
	\label{fig:ELC}
\end{figure*}

\label{sec:ELC}
In this section, the EMA linear combination (ELC) algorithm is presented and analyzed. This algorithm is composed of two main parts: 
$\alpha$ and $\lambda$ selection ($\lambda$-s) based on $\mathcal{D}_{tr}$, 
and evaluation based on $\mathcal{D}_{te}$. 
Steps are summarized in the example of Fig.~\ref{fig:ELC}, 
which is meant to clarify ELC operation.

\subsection{$\lambda$-s: Initial selection of $\alpha$ weights}
\label{sub:selection}
In the first step of the algorithm, a number of EMA models were executed concurrently with different values of the $\alpha$ weight,
taken from a \textit{starting sequence} $\alpha\mathrm{S}$ (i.e., an ordered sequence of numbers) defined as
\begin{equation}
\label{eq:alphaS}
\alpha\mathrm{S}=( \alpha_n \mid \alpha_n=\alpha^* \cdot r^{n}),
\end{equation}
where $\alpha_n=\alpha^* \cdot r^{n}$ is sort of a 
finite-size \textit{geometric progression} 
with common ratio $r$
in which $n$ can also assume negative values. 
In particular, $n \in [-N_l, -N_l+1,...,-1, 0, 1, ..., N_u-1, N_u]$, 
where $N_l$ and $N_u$ are used to specify the lower and upper bound for the selection of $\alpha$ weights, respectively.
The length of the sequence $\alpha\mathrm{S}$ is $|\alpha\mathrm{S}|=N_l+1+N_u$. 
As an example, when $r=2$, $N_l=2$, and $N_u=4$, the starting sequence 
includes $7$ values and
is $\alpha\mathrm{S}=( \frac{\alpha^*}{4}, \frac{\alpha^*}{2}, \alpha^*, 2\alpha^*, 4\alpha^*, 8\alpha^*, 16\alpha^* )$.

Since the purpose of the \mbox{$\lambda$-s} algorithm is to select the set of $\alpha$ weights that are more suitable for a specific database, 
the space of $\alpha$ weights should be sampled in such a way not to discard any relevant values. 
Using a geometrical progression permits to have more $\alpha$ weights near $\alpha^*$, progressively making them less frequent when the distance from $\alpha^*$ increases. 
This selection strategy 
was verified experimentally and compared to a uniform sampling of the $\alpha$ space. 
Results, not reported here for space reasons,
confirm that this choice was the most appropriate.

Decreasing $r$ permits to analyze the space of $\alpha$ weights with finer granularity, but execution time grows because the number of EMA models to be checked is higher
(given the lower and upper bounds). 
$N_l$ and $N_u$ were selected large enough to embrace values of $\alpha$ that are three orders of magnitude smaller and larger than $\alpha^*$, respectively.

Let $(y^{(\alpha_1)}_i, y^{(\alpha_2)}_i,...,y^{(\alpha_{|\alpha\mathrm{S}|})}_i)$ be the sequence of predictions obtained with the EMA models parameterized according to the values in sequence $\alpha\mathrm{S}$. 
For the example above, $\alpha_1=\frac{\alpha^*}{4}, \alpha_2=\frac{\alpha^*}{2},\alpha_3=\alpha^*,...,\alpha_{|\alpha\mathrm{S}|}=16\alpha^*$.

\subsection{$\lambda$-s: Linear combination}
To improve prediction accuracy, a linear combination (COM) can be evaluated from the output values $y^{(\alpha_j)}_i$ produced by several EMA models,
each one characterized by weight $\alpha_j$,
\begin{eqnarray}
    \label{eq:com_model}
    y^{\mathrm{COM}}_i = \sum_{j=1}^{|\alpha\mathrm{A}|} \lambda^{(\alpha_j)} \cdot y_i^{(\alpha_j)},  
\end{eqnarray}
where coefficients $\lambda^{(\alpha_j)}$ are selected so that
\begin{eqnarray}
    \label{eq:com_constraints}
    \sum_{j=1}^{|\alpha\mathrm{A}|} \lambda^{(\alpha_j)} = 1,
\end{eqnarray}
and 
$\alpha\mathrm{A}=(\alpha_1, \alpha_2,...,\alpha_{|\alpha\mathrm{A}|})$
%is 
represents in a generic way the sequence of $\alpha$ weights 
of the considered EMA models.
Index $j$ in \eqref{eq:com_model} selects both 
the specific EMA model (characterized by the value of weight $\alpha_j$, which determines the associated prediction $y_i^{(\alpha_j)}$)
and the coefficient $\lambda^{(\alpha_j)}$ used to weight its contribution in the linear combination.

The COM model in \eqref{eq:com_model} coincides with a multipole low-pass IIR filter 
characterized by the transfer function
\begin{align}
    H^{(\alpha\mathrm{A})}(z) &= 
    \sum_{j=1}^{|\alpha\mathrm{A}|} \lambda^{(\alpha_j)} \cdot H^{(\alpha_i)}(z). 
\end{align}

As analysed in a very preliminary way in \cite{WFCS2023}, the linear combination of EMA models may offer better prediction accuracy 
than any one of them considered separately. 
In that paper, the sequence of $\alpha$ weights was statically selected as 
$\alpha\mathrm{A}=(\frac{\alpha^*}{3}, \alpha^*, 3\alpha^*)$ and the three models were equally weighted ($\lambda^{(\alpha_1)}=\lambda^{(\alpha_2)}=\lambda^{(\alpha_3)}=\frac{1}{3}$). 
In this paper that solution is tangibly enhanced thanks to the ELC algorithm, which permits the automatic selection of both the $\alpha_j$ weights and $\lambda^{(\alpha_j)}$ coefficients of the linear combination by means of a suitable training phase
performed on a relevant database.

\subsection{$\lambda$-s: Minimizing the MSE}
\label{sub:minimization}
Among the potentially many $\alpha$ values included in the starting sequence $\alpha\mathrm{S}$, we aim to assess which ones affect prediction accuracy more.
The selection of a small subset of $\alpha\mathrm{S}$, 
which includes the most relevant EMA weights,
has been addressed as a minimization problem of the mean squared error $\mu_{e^2}$ of the COM model (\ref{eq:com_model}), constrained by (\ref{eq:com_constraints}) and with the  boundaries
\begin{eqnarray}
    \label{eq:com_boundaries}
    0 \le \lambda^{(\alpha_j)} \le 1,
\end{eqnarray}
over the training database $\mathcal{D}_{tr}$.

The method chosen to perform the minimization was the L-BFGS-B \cite{L-BFGS}, a quasi-Newton optimization algorithm. 
It is a gradient-based optimization algorithm that iteratively updates an estimate of the solution using gradient and curvature information, while also handling bound constraints. It is designed for high-dimensional problems and can efficiently find the minimum of a function.

The outcome of the MSE minimization problem is a sequence 
$\mathrm{\lambda\mathrm{S}}=(\lambda^{(\alpha_1)}, \lambda^{(\alpha_2)},...,\lambda^{(\alpha_{|\alpha\mathrm{S}|})})$
of optimal coefficients for the linear combination,
each one associated to a specific value in the sequence $\mathrm{\alpha\mathrm{S}}$
(by definition, the size of sequences $\alpha\mathrm{S}$ and $\lambda\mathrm{S}$ is the same, i.e., $|\alpha\mathrm{S}|=|\lambda\mathrm{S}|$).

\subsection{$\lambda$-s: Final selection of $\alpha$ weights}
We observed that part of the combination coefficients evaluated by the minimization algorithm in the previous Subsection~\ref{sub:minimization} were rather small. 
Consequently, their contribution to the final prediction is irrelevant.
This is even more true if they are considered on a database different from the one used to perform minimization, e.g., $\mathcal{D}_{te}$. 
In addition, if the $\alpha\mathrm{S}$ and $\lambda\mathrm{S}$ sequences are shortened, 
the algorithm becomes faster and more suitable to be implemented in embedded devices
with scarce computational power.

Starting from $\alpha\mathrm{S}$, a new \textit{final sequence} $\alpha\mathrm{F}$ is obtained by selecting part of the values $\alpha_i$ in $\alpha\mathrm{S}$ for which the corresponding $\lambda^{(\alpha_i)}$ combination coefficients are larger.
Practically, 
let $\overline{\alpha\mathrm{S}}
=(\overline\alpha_1, \overline\alpha_2,...,\overline\alpha_{|\alpha\mathrm{S}|})$
be a version of the sequence $\alpha\mathrm{S}$ 
sorted in decreasing order of the associated value of $\lambda^{(\overline\alpha_i)}$.
Let $N_\alpha=|\alpha\mathrm{F}|$ represent the number of EMA models that are 
eventually selected for testing.
It can be evaluated as
\begin{equation}
\label{eq:selection1}
    N_\alpha = 
    \min \biggl( i \in 1...|\overline{\alpha\mathrm{S}}| 
    \mid \sum_{j=1}^{i} 
    {\lambda^{(\overline\alpha_j)}} \ge \lambda_{\mathrm{max}} \biggr),
\end{equation}
where $\lambda_\mathrm{max}$ is a threshold used to discriminate which 
weights must be included in $\alpha\mathrm{F}$.
Sequence $\alpha\mathrm{F}$ can then be derived as
\begin{equation}
\label{eq:selection2}
\alpha\mathrm{F} = \left( \overline\alpha_i \in \overline{\alpha\mathrm{S}} 
    \mid i \leq N_\alpha
    \right).
\end{equation}
For instance, setting $\lambda_\mathrm{max}=0.9$  means to take 
the minimum number of elements from $\alpha\mathrm{S}$ 
so that the sum of the related coefficients in $\lambda\mathrm{S}$ exceeds $90\%$.
This is the same as keeping only the EMA filters that mostly contributed to reducing the MSE ($\mu_{e^2}$).

The minimization algorithm is then applied a second time to the sequence $\alpha\mathrm{F}$, 
to obtain the new  $\lambda^{(\alpha_i)}$ coefficients that minimize the MSE for $\mathcal{D}_{tr}$. 
These new $\lambda^{(\alpha_j)}$ values constitute the actual final sequence $\lambda\mathrm{F}$,
which will be used for testing.
Setting $\lambda_\mathrm{max}=1.0$ implies that the second minimization of the linear combination is not performed.

\subsection{Evaluation}
Given the new final sequence $\alpha\mathrm{F}=(\alpha_1, \alpha_2,...,\alpha_{|\alpha\mathrm{F}|})$  and $\lambda\mathrm{F}=(\lambda^{(\alpha_1)}, \lambda^{(\alpha_2)},...,\lambda^{(\alpha_{|\lambda\mathrm{F}|})})$, the linear combination in (\ref{eq:com_model}) provides the prediction $y^{\mathrm{COM}}_i$ of the COM model. 
The primary outcome of this work is to determine if the $\alpha$ weights and the $\lambda^{(\alpha_j)}$ coefficients
of the linear combination, which are optimal for $\mathcal{D}_{tr}$ by considering $\mu_{e^2}$ as the error metric, are also optimal for $\mathcal{D}_{te}$ databases,
by considering $\mu_{e^2}$ and, possibly, other error functions like $\mu_{e}$.

\section{Results}
\label{sec:results}
\subsection{Database acquisition}
\label{sub:DB}
The acquisition of $\mathcal{D}_{tr}$ and $\mathcal{D}_{te}$ was performed with the same experimental setup used in \cite{WFCS2023}, which consists of four Wi-Fi adapters of type TP-Link TL-WDN4800 that are configured to transmit periodically (with a frequency of $\unit[2]{Hz}$) frames of dimension $\unit[50]{B}$ directed to four different access points. 
The arrival of the ACK frame associated with every data frame is used to determine its transmission outcome (success or failure), which is then stored in a database.

Coherently with the rules of the software-defined MAC (SDMAC) framework \cite{8477080,7991945}, the device driver was modified in order to: disable automatic retransmissions, disable frame aggregation, disable the backoff procedure, and fix the transmission speed to $\unit[54]{Mb/s}$.
These changes permitted us to sample the channel at a fixed rate, 
preventing the effects of some mechanisms in Wi-Fi that could distort measurements, for example by automatically retransmitting frames when the ACK does not arrive back to the sender node.

Training and test databases were acquired on four different channels, 
and their size is considerably larger than
prior works on this topic like \cite{WFCS2023}.
Acquisition  spanned over $21.2$ days for $\mathcal{D}_{tr}$ and $12.8$ days for $\mathcal{D}_{te}$ on channel $1$ ($\unit[2.412]{GHz}$), 
$21.9$ days for $\mathcal{D}_{tr}$ and $10.6$ days for $\mathcal{D}_{te}$ on channel $5$ ($\unit[2.432]{GHz}$), 
$24.9$ days for $\mathcal{D}_{tr}$ and $15.2$ days for $\mathcal{D}_{te}$ on channel $9$ ($\unit[2.452]{GHz}$), 
and $25.5$ days for $\mathcal{D}_{tr}$ and $15.2$ days for $\mathcal{D}_{te}$ on channel $13$ ($\unit[2.472]{GHz}$).

\subsection{EMA baseline}
\label{sub:EMA_baseline}
\begin{figure}[t]
    \begin{center}
    \includegraphics[width=0.9\columnwidth]{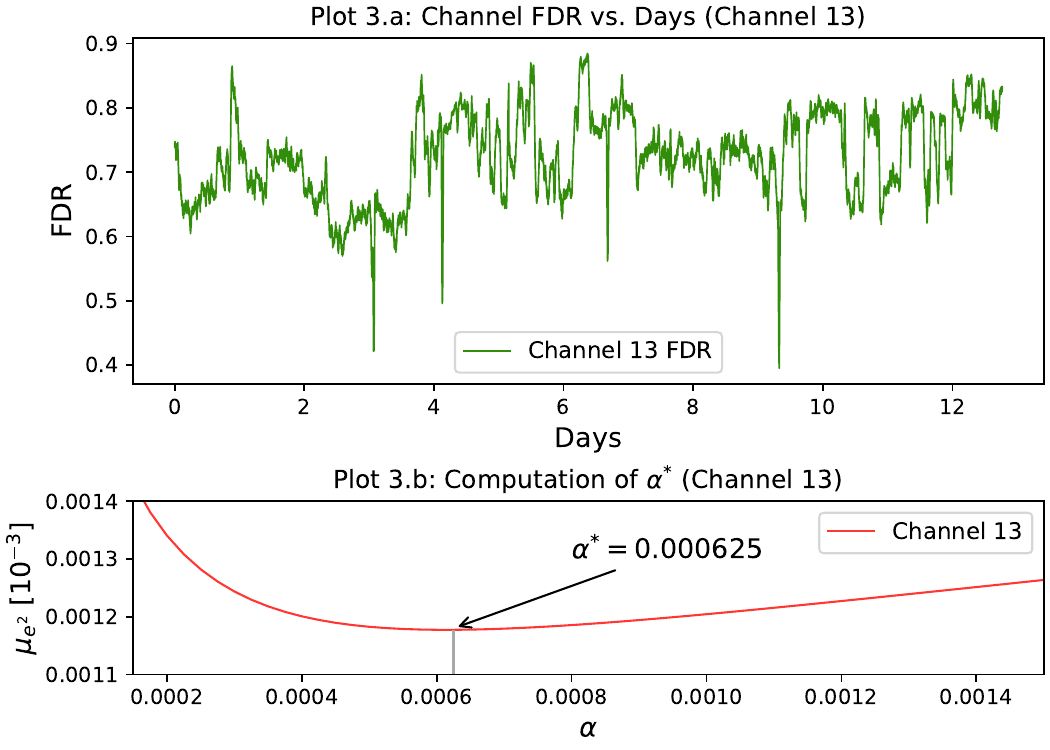}
    \end{center}
    \caption{Example case of channel $13$: timing diagram for FDR in $\mathcal{D}_{te}$ (Plot~\ref{fig:plot}.a), and computation of $\alpha^*$ from the MSE in $\mathcal{D}_{tr}$ (Plot~\ref{fig:plot}.b).}
	\label{fig:plot}
\end{figure}

\begin{table*}[t]
  \caption{Comparison between EMA and ELC for the considered Wi-Fi channels based on statistics about prediction errors.}
  \label{tab:res_global}
  \footnotesize
  \begin{center}
    \tabcolsep=0.18cm
    \def\arraystretch{1.25}
    \begin{tabular}{ccc|ccc|cccccc|cccc}
    Channel & Model & Model  & $\mu_{e^2}$ & $e^2_{\mathrm{p}_{95}}$ & $e^2_{\mathrm{max}}$ & $\mu_{|e|}$ & $\sigma_{|e|}$ & ${|e|}_{\mathrm{p}_{90}}$ & ${|e|}_{\mathrm{p}_{95}}$ & ${|e|}_{\mathrm{p}_{99}}$ & ${|e|}_{\mathrm{max}}$ & ${e}_{\mathrm{min}}$ & ${e}_{\mathrm{p}_{5}}$ & ${e}_{\mathrm{p}_{95}}$ & ${e}_{\mathrm{max}}$ \\
    & & parameters & \multicolumn{3}{c|}{$[\cdot 10^{-3}]$} & \multicolumn{6}{c|}{[\%]} & \multicolumn{4}{c}{[\%]} \\
    \hline
    \multirow{2}{*}{$1$} 
        & EMA & $\alpha^*=0.001$ & 2.62 & 13.88 & 101.51 & 3.38 & 3.85 & 8.64 & 11.78 & 16.93 & 31.86 & -24.75 & -8.16 & 9.14 & 31.86  \\
        & ELC& $N_{\alpha}=5$ & 2.12 & 11.09 & 73.49 & 3.13 & 3.38 & 7.80 & 10.53 & 15.04 & 27.11 & -22.97 & -7.81 & 7.78 & 27.11\\
    \hline
    \multirow{2}{*}{$5$} 
        & EMA &  $\alpha^*=0.000075$ & 1.07 & 2.71 & 74.61 & 2.17 & 2.44 & 4.31 & 5.20 & 10.73 & 27.31 & -27.31 & -4.56 & 4.12 & 7.58 \\
        & ELC & $N_{\alpha}=3$ & 0.95 & 2.39 & 75.49 & 1.98 & 2.37 & 3.84 & 4.89 & 11.06 & 27.48 & -27.48 & -4.13 & 3.59 & 16.14\\
    \hline
    \multirow{2}{*}{$9$} 
        & EMA &  $\alpha^*=0.000075$ & 8.62 & 42.60 & 363.64 & 5.77 & 7.27 & 13.61 & 20.64 & 34.91 & 60.30 & -60.30 & -20.51 & 11.21 & 21.01  \\
        & ELC & $N_{\alpha}=2$ & 6.95 & 36.60 & 376.06 & 5.08 & 6.61 & 11.94 & 19.13 & 33.28 & 61.32 & -61.32 & -17.91 & 9.62 & 33.32 \\
    \hline
    \multirow{2}{*}{$13$} 
        & EMA & $\alpha^*=0.000625$ & 0.72 & 3.19 & 19.75 & 1.94 & 1.85 & 4.31 & 5.65 & 8.81 & 14.05 & -12.07 & -4.28 & 4.35 & 14.05 \\
        & ELC& $N_{\alpha}=4$ & 0.69 & 2.88 & 22.60 & 1.95 & 1.77 & 4.16 & 5.37 & 8.53 & 15.03 & -11.76 & -4.00 & 4.35 & 15.03 \\
    \hline
    \end{tabular}
    \end{center}
\end{table*}

Plot~\ref{fig:plot}.a reports the evolution of FDR for $\mathcal{D}_{te}$ referred to channel $13$, computed with (\ref{eq:target}) and $N_f=3600$. 
As can be seen, the quality of this Wi-Fi channel experienced consistent variations (from $60\%$ to $80\%$), and sometimes disturbance spikes appeared suddenly. 
Channel $5$ is characterized by a similar pattern, while channel $1$ and, especially, channel $9$ are characterized by faster and larger FDR variations (from $40\%$ to $90\%$).

As highlighted in Plot~\ref{fig:plot}.b, equation (\ref{eq:ema_opt}) is exploited to find the optimal values of $\alpha$ weights (that is, parameters $\alpha^*$) for the four considered channels, as reported in the third column of Table~\ref{tab:res_global} for rows labeled ``EMA''.
Baseline results, obtained with the plain EMA model using these $\alpha^*$ values, are also reported in the table.

As expected, channels $13$ and $5$ show lower prediction errors, because they are characterized by smoother FDR variations. 
In these cases, the FDR can be predicted with an average absolute error $\mu_{|e|}$ of about $2\%$, and also high-order percentiles like ${|e|}_{p99}$ are bounded to prediction errors smaller than $\sim 10\%$.
On the contrary, due to abrupt and difficult-to-predict FDR variations, the performance of channel $9$ is rather poor, 
i.e., $\mu_{|e|}=5.77\%$ and ${|e|}_{p99}=34.91\%$.

Above considerations also hold for the other indicators about prediction accuracy, which are related to the real ($e$) and absolute ($|e|$) errors.

\subsection{ELC}
\label{sub:ELC}

\begin{figure}[t]
    \begin{center}
    \includegraphics[width=0.9\columnwidth]{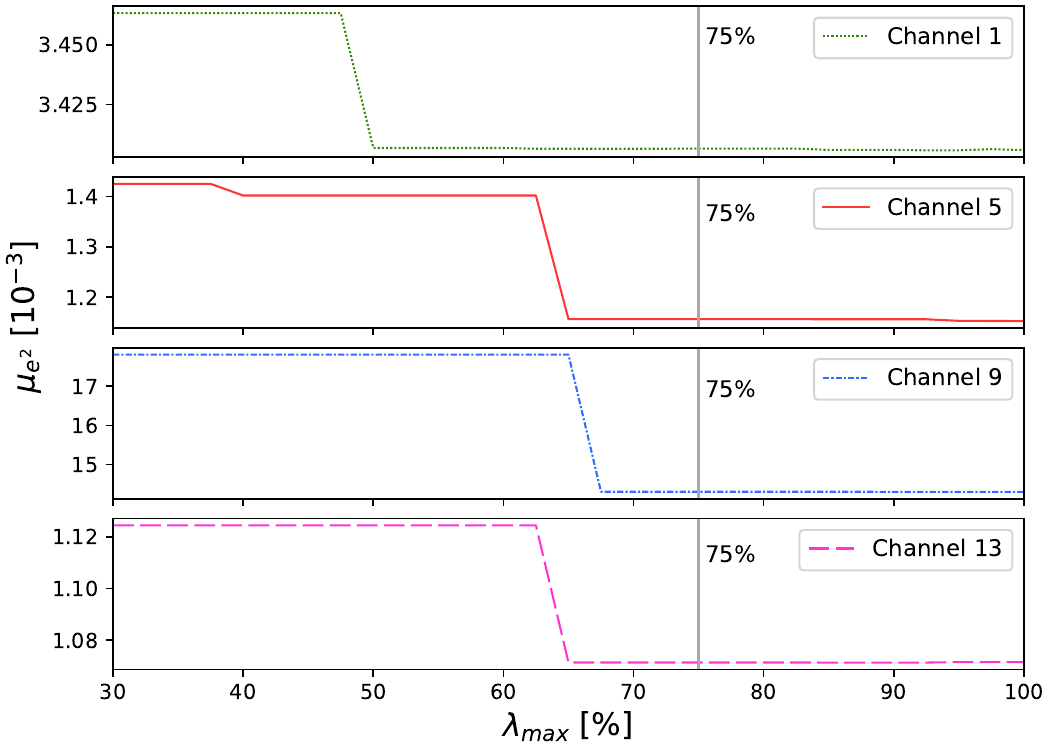}
    \end{center}
    \caption{MSE ($\mu_{e^2}$) vs. $\lambda_{\mathrm{max}}$ 
    for the considered Wi-Fi channels
    (the value $\lambda_{\mathrm{max}}=75\%$ was selected).}
	\label{fig:error}
\end{figure}
The parameters used as the initial selection of $\alpha$ weights 
for the evaluation of the ELC model (Subsection~\ref{sub:selection}) 
are $N_l=17$, $N_u=17$, and $r=1.5$. 
Consequently, the starting sequence $\alpha\mathrm{S}$ included $|\alpha\mathrm{S}|=35$ $\alpha$ weights.

After the minimization step described in Subsection~\ref{sub:minimization}, 
the sorted sequence of $\lambda$ (i.e., $\overline{\lambda\mathrm{S}}$) was used, 
in conjunction with (\ref{eq:selection1}) and (\ref{eq:selection2}), 
to select the most significant coefficients $\lambda^{(\overline\alpha_j)}$ 
by varying the $\lambda_{\mathrm{max}}$ parameter. 

The plots in Fig.~\ref{fig:error} report the mean squared error $\mu_{e^2}$ obtained with different values of $\lambda_{\mathrm{max}}$ for the four training databases $\mathcal{D}_{tr}$ related to channels $1$, $5$, $9$, and $13$. 
As can be seen, when $\lambda_{\mathrm{max}} > 75\%$ there are practically no improvements regarding $\mu_{e^2}$. For this reason, the value $\lambda_{\mathrm{max}}=75\%$ has been selected for the entire experimental campaign.

Rows labeled ``ELC'' in Table~\ref{tab:res_global} show the results obtained for the four test databases $\mathcal{D}_{te}$ that refer to channels $1$, $5$, $9$, and $1$3, after all the steps of the ELC algorithm have been performed as described in Section~\ref{sec:ELC}.
The improvements this approach provides over plain EMA 
(a single optimized exponential moving average) are noticeable in all the four experimental conditions. 

Results in Table~\ref{tab:res_global} show that, 
excluding some of the maximum error values (e.g., $e^2_{\mathrm{max}}$ and ${|e|}_{\mathrm{max}}$ for channels $5$, 9, and 13) 
and some high-order percentiles (e.g., ${|e|}_{\mathrm{p99}}$ for channels $5$ and $9$), 
the ELC model practically always behaves better than the EMA model. 
In particular, the relative reduction of $\mu_{e^2}$, which is the objective function minimized in this work, is quite consistent, and corresponds to $19.1\%$, $11.2\%$, $19.4\%$, and $4.2\%$ for channels 1, $5$, $9$, and $13$, respectively. 

Regarding $\mu_{|e|}$,
with the only exception of channel $13$ (where the error remains quite stable), 
the relative reduction of the absolute error is $7.4\%$, $8.8\%$, and $12.0\%$ for channels $1$, $5$, and $9$, respectively. 
All these results confirm the higher prediction accuracy of the ELC-based model compared to EMA. 
Again, improvements concerning $\mu_{e^2}$ are better than those related to $\mu_{|\mu_e|}$, 
since the $\lambda$-s algorithm 
performs minimization on the MSE. %$\mu_{e^2}$.

Finally, it is worth noting that the number $N_{\alpha}$ of EMA models that were selected by the $\lambda$-s algorithm slightly depends on the specific channel, 
but remains quite small when compared to the cardinality of the initial set of EMA models ($|\alpha\mathrm{S}|=35$). 
This means that ELC is not very demanding from the point of view of computation resources.

\section{Conclusions}
\label{sec:conclusions}
The ability to reliably predict the channel quality is the most important requirement for implementing wireless network protocols that proactively adapt their behavior to the characteristics of the communication medium. 
In this paper, a new method named ELC was presented and evaluated. 
By linearly combining EMA filters configured with different parameters, 
it permits to improve the accuracy with which the future quality of a wireless channel, 
expressed in terms of frame delivery ratio, is predicted.

The capabilities of the proposed method were experimentally assessed on a set of databases acquired from a real \mbox{Wi-Fi} installation, and a substantial reduction 
of the prediction error was observed with respect to the EMA model in every condition we analyzed. 
Moreover, the low computational complexity of ELC, compared to methods based on artificial neural network, 
makes it suitable for adoption in low-cost low-power devices where communication takes typically place on air, 
e.g., motes in wireless sensor networks (WSN).

Future research directions include the analysis of ELC in different experimental conditions, the use of ANNs for selecting frame-by-frame the best EMA model, 
and the use of ANNs for non-linear combinations of EMA models.
\bibliographystyle{IEEEtran}
\bibliography{bibliography}

\end{document}